# Fluid-structure interaction: Insights into biomechanical implications of endograft after thoracic endovascular aortic repair

Yonghui Qiao [a], Le Mao [b], Ying Ding [c], Ting Zhu [b], Kun Luo [a,]* and Jianren Fan [a,]*

[a] *State Key Laboratory of Clean Energy Utilization, Zhejiang University, Hangzhou, China*

[b] *Department of Vascular Surgery, Zhongshan Hospital, Fudan University, Shanghai, China*

[c] *Department of Radiology, Zhongshan Hospital, Fudan University, Shanghai, China*

\* Corresponding author:

**Kun Luo**

State Key Laboratory of Clean Energy Utilization, Zhejiang University

38 Zheda Road, Hangzhou, China, 310027

Phone: +86-18667160183    E-mail：zjulk@zju.edu.cn

**Jianren Fan**

State Key Laboratory of Clean Energy Utilization, Zhejiang University

38 Zheda Road, Hangzhou, China, 310027

Phone: +86-13336107178    E-mail：fanjr@zju.edu.cn

**Abstract:** Thoracic endovascular aortic repair (TEVAR) has developed to be the most effective treatment for aortic diseases. This study aims to evaluate the biomechanical implications of the implanted endograft after TEVAR. We present a novel image-based, patient-specific, fluid-structure computational framework. The geometries of blood, endograft, and aortic wall were reconstructed based on clinical images. Patient-specific measurement data was collected to determine the parameters of the three-element Windkessel. We designed three postoperative scenarios with rigid wall assumption, blood-wall interaction, blood-endograft-wall interplay, respectively, where a two-way fluid-structure interaction (FSI) method was applied to predict the deformation of the composite stent-wall. Computational results were validated with Doppler ultrasound data. Results show that the rigid wall assumption fails to predict the waveforms of blood outflow and energy loss (EL). The complete storage and release process of blood flow energy, which consists of four phases is captured by the FSI method. The endograft implantation would weaken the buffer function of the aorta and reduce mean EL by 19.1%. The closed curve area of wall pressure and aortic volume could indicate the EL caused by the interaction between blood flow and wall deformation, which accounts for 68.8% of the total EL. Both the FSI and endograft have a slight effect on wall shear stress-related-indices. The deformability of the composite stent-wall region is remarkably limited by the endograft. Our results highlight the importance of considering the interaction between blood flow, the implanted endograft, and the aortic wall to acquire physiologically accurate hemodynamics in post-TEVAR computational studies and the deformation of the aortic wall is responsible for the major EL of the blood flow.

# 1. Introduction

Thoracic endovascular aortic repair (TEVAR) has emerged as the predominant method with low morbidity and mortality in the treatment of patients with aortic diseases [1, 2]. Most patients experience significant improvements after TEVAR. However, there are some potential clinical risks associated with implanted endograft, such as endoleak, stent-graft-induced new entries, aortic wall rupture, and endograft migration [3]. Therefore, there is an urgent demand for bridging the biomechanical implications of the implanted endograft with the complications of TEVAR.

Recently, several researchers have primarily explored the clinical negative outcome of endograft. Kanaoka et al. [4] evaluated risk factors for early Type I endoleak after TEVAR for aneurysm, while the natural history of endoleak remains unclear. Dong et al. [5] reported retrograde type A aortic dissection after endovascular endograft placement for treatment of type B dissection and they also found that stress-induced injury from the endograft should be concerned during TEVAR [6]. Daye et al. [7] summarized the common endograft-related complications of TEVAR in the treatment of aneurysms and discussed re-intervention strategies. Endograft migration was also evaluated through retrospective analysis, where aortic elongation and thoracic aortic aneurysm were found to be risk factors [8]. The effect of the implanted endograft is getting more and more attention in the clinic.

Image-based computational fluid dynamics has provided physical insights into the clinical efficacy of TEVAR. However, most previous postoperative computational hemodynamic studies have neglected the effect of the implanted endograft. Figueroa et al. [9] firstly assessed pulsatile displacement forces acting on realistic thoracic aortic endograft and discussed the risk of endograft migration. Later, they developed a computational framework for investigating the positional stability of aortic endograft [10]. Menichini et al. [11] found that high stent-graft tortuosity could result in high wall stress and may induce new aortic entries. van Bakel et al. [12] reported that stiffness mismatch between the stent-graft and the aorta likely leads to the rupture of the ascending aorta. We previously discovered that the configurations of the complex stent-grafts have a significant impact on postoperative aortic hemodynamics [13]. Nevertheless, the understanding of the interplay between the blood flow, endograft, and aortic wall is still insufficient and the role of the endograft requires further investigation.

The present study aims to quantitatively elucidate the complex biomechanical implication of the endograft by taking the interactive coupling of the blood flow, endograft, and aortic wall into account. In this study, we developed a novel image-based, patient-specific, fluid-structure computational framework to investigate the biomechanical effect of the endograft noninvasively. Three scenarios with rigid wall assumption, blood-wall interaction, blood-endograft-wall interplay are constructed basing on a clinical patient treated by TEVAR. The crucial hemodynamics and the postoperative deformation of the aortic wall are quantitatively explored to reveal the biomechanical implication of the implanted endograft.

## 2. Methodologies

### *2.1 Patient information*

The patient in this study was a 49-year-old male who had an aortic arch aneurysm involving the left subclavian artery (LSA). Informed consent for inclusion in the present study was obtained from the patient. Considering the insufficient proximal landing zone, the LSA was intentionally covered by the Captivia thoracic stent-graft (Medtronic Vascular, Santa Rosa, CA, USA) during TEVAR surgery. To reduce the risk of complications related to the coverage of the LSA, *in situ* fenestration technique [14] was applied to revascularize the blocked LSA with a Viabahn covered stent (WL Gore & Associates, Flagstaff, Ariz, USA). The cardiac output and blood flow distribution during a cardiac cycle was acquired by using Doppler ultrasound measurement and phase-contrast magnetic resonance imaging (PC-MRI).

## *2.2 Geometries and meshes*

Clinical computed tomography angiography (CTA) images were used to segment and reconstruct the 3D postoperative aortic geometry with three supra-aortic branches reserved by using Mimics 19.0 (Materialise, Leuven, Belgium). The locations of stent-graft implanted in LSA and endograft were also acquired from CTA data. The geometric model of the stent-graft can be directly identified by the gap around the LSA. Considering that CTA cannot accurately capture the complete endograft structure, the thickness of the endograft was assumed to be 0.6 mm according to the medical device supplier. The aortic wall was simplified to be a uniform thickness of 2 mm which was generated by extending the inner surface of the aortic wall outward. The inner surface of the aortic wall was the combination of the outer surfaces of the blood flow domain and two implanted stent-grafts. Fig. 1 shows the geometric models of the blood flow domain, implanted endograft, and aortic wall. Three postoperative scenarios with rigid wall assumption, blood-wall interaction, blood-endograft-wall interplay, respectively, were designed to evaluate the biomechanical implications of the endograft (Fig. 2). It should be emphasized that the stent-graft implanted in the LSA was also included in the two fluid-structure interaction (FSI) scenarios.

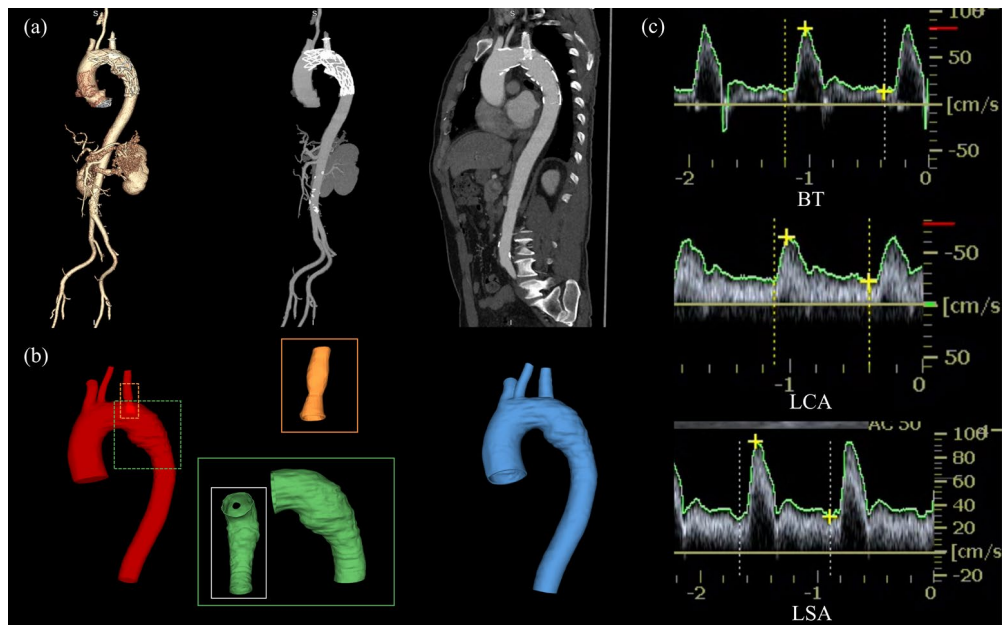

**Fig. 1.** Geometric model and boundary condition. (a) CTA scan images. (b) Reconstructed 3D geometric models: aorta (red), LSA stent-graft (orange), endograft (green) and vessel wall (blue). (c) Blood flow rate from Doppler ultrasound measurement (BT: brachiocephalic trunk; LCA: left carotid artery; LSA: left subclavian artery).

In the present study, all the computational meshes were generated by using ANSYS-ICEM 16.1 (ANSYS Inc, Canonsburg, USA). The blood flow domain had more than 1,500,000 unstructured elements consisted of tetrahedral meshes in the flow center region and eight prism layers near the wall. The LSA stent-graft and endograft were discretized into 2,000 and 50,000 unstructured tetrahedral elements, respectively. The two geometric models of the aortic wall in FSI scenarios both had approximately 100,000 elements. Mesh independent tests were performed and the differences in wall shear stress (WSS) and wall displacement between the chosen meshes and finer meshes were less than 3%.

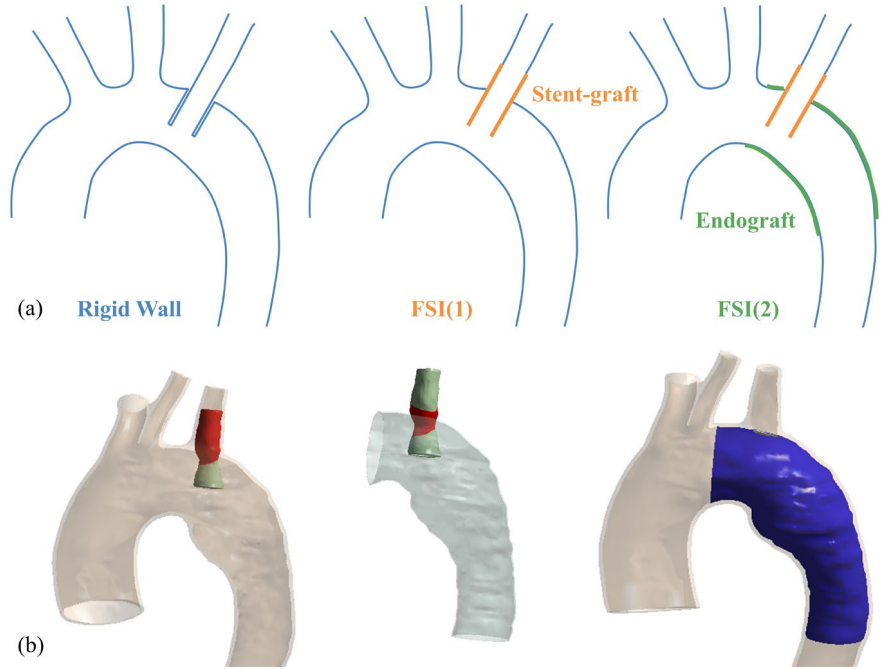

**Fig. 2.** Diagram of the three scenarios of the postoperative aorta. (a) Rigid wall, Fluid-structure interaction, and Fluid-structure interaction with endograft. (b) Complex three-body contact interactions between the LSA stent-graft, endograft, and aortic wall.

## *2.3 Numerical model and boundary conditions*

The blood was considered as a non-Newtonian fluid with the Carreau-Yasuda viscosity model and a density of 1080 kg/m$^3$ [15]. The LSA stent-graft and aortic endograft were both assumed to be linear elastic material with a Young's modulus of 8.4 MPa and a Poisson ratio of 0.3 [16]. A similar modeling strategy was applied to the aortic wall, where Young's modulus was relatively low (2 MPa) [17] and the Poisson ratio was set to 0.45 [18].

A normalized physiological flow waveform from a previous study [19] was tailored by shift-and-scaling to the patient-specific cardiac cycle and output (Fig. 3a). In the present study, the maximum Reynolds numbers ($Re_{max}$) Womersley numbers ($\alpha$) are 4792 and 26.2, respectively at peak systole. The maximum Reynolds number ($Re_{max}$) is lower than the threshold range of $Re_c = k\alpha$ when the coefficient $k$ takes the minimum value (250) [20]. Therefore, the blood flow was assumed to be laminar.

The three-element Windkessel model was coupled to account for the flow-pressure relationship in the outlet boundaries. To determine the three parameters of the Windkessel model, Pirola et al. [21] proposed a workflow, which needs two input parameters. One is the blood flow rate during a cardiac cycle, which was derived from Doppler ultrasound measurement in this study (Fig. 1). The blood flow calculation process of the supra-aortic branches is as follows:

$$S = \sum_{1}^{N}(v_n - v_{n-1})(t_n - t_{n-1}) \tag{1}$$

$$\bar{v} = S/T - v_0 \tag{2}$$

$$Q = \bar{v}/2 \times A \times HR \tag{3}$$

where N is the total number of measuring points in Fig. 1. $\bar{v}$ is the mean velocity, T refers to a cardiac cycle, and $v_0$ is the reference velocity. HR indicates the heart rate. The other parameter is the average blood pressure, where the slight difference between different aortic outlets was neglected for the sake of simplicity. The systolic and diastolic pressures of the left brachial artery were measured by an arm cuff sphygmomanometer to calculate the average aortic pressure. Parameters of the three-element Windkessel model are shown in Table 1 and the same set of model parameters was applied in the three computational scenarios.

## 2.5 Fluid-structure interaction

In this paper, a two-way FSI method was used to capture the deformations of the implanted stent-grafts and the aortic wall. The center points of boundary surfaces of the stent-grafts and aortic wall were fixed to achieve numerical stability, where expansion and contraction were still allowed for above solid structures [22, 23]. We preloaded the measured diastolic pressure on the outer surface of the aortic wall to simulate the external forces of the surrounding tissues [22]. A Rayleigh damping was applied to further improve the convergence of the two FSI scenarios [24, 25]. The contact interactions between the LSA stent-graft, endograft, and the aortic wall were accounted for by a basic Coulomb frictional model and the coefficient of friction was 0.3.[10]

$$\tau_{lim} = \mu P \qquad (4)$$

where $\tau_{lim}$ is limit frictional stress. $\mu$ and $P$ refer to friction coefficient and contact normal pressure, respectively. In general, the frictional stress is lower than the limit, the contact cohesion between different surfaces is stable and relative sliding is restricted.

**TABLE 1.** Parameters of the three-element Windkessel model.

| OUTLET | $R_1$ [$10^7$ Pa s m$^{-3}$] | $C$ [$10^{-10}$ m$^3$ Pa$^{-1}$] | $R_2$ [$10^8$ Pa s m$^{-3}$] |
|---|---|---|---|
| BT | 5.368 | 16.14 | 10.55 |
| LCA | 26.05 | 3.450 | 49.27 |
| LSA | 17.81 | 9.384 | 17.29 |
| DA | 1.286 | 101.7 | 1.631 |

$R_1$: proximal resistance; $R_2$: distal resistance; $C$: vessel compliance; BT: brachiocephalic artery; LCA: left carotid artery; LSA: left subclavian artery; DA: descending aorta.

All the simulations were carried out on ANSYS Workbench 16.1 (ANSYS Inc, Canonsburg, USA), where Transient Structural was coupled with CFX to implement the complex FSI. A constant time step of 5 ms was adopted considering computational efficiency. Five cardiac cycles were required before simulations reach a periodic solution and we present the results of the fifth cardiac cycle.

*2.6 Hemodynamic parameters*

Energy Loss (EL) of the blood flow and WSS-related indices were evaluated in this study. The energy difference between the inlet and outlets of the aorta indicates the EL during a cardiac cycle. The formula of EL is defined as follows [26]:

$$\text{EL} = \sum\nolimits_{\text{Inlet}} (TP*Q) - \sum\nolimits_{\text{Outlets}} (TP*Q) \quad (5)$$

where $Q$ is the blood flow rate. TP indicates the total pressure.

$$TP = \frac{1}{2}\rho |\vec{u}|^2 + P \quad (6)$$

where $u$ and $P$ refer to velocity and pressure, respectively.

Compared with instantaneous value, WSS-related indices averaged by a cardiac cycle are more reliable. We calculated the time-averaged WSS (TAWSS) and oscillatory shear index (OSI). The corresponding formulas are as follows:

$$\text{TAWSS} = \frac{1}{T}\int_0^T |\tau(t)| dt \quad (7)$$

$$\text{OSI} = 0.5\left(1 - \frac{\left|1/T \int_0^T \tau(t) dt\right|}{\text{TAWSS}}\right) \quad (8)$$

where $\tau(t)$ is the WSS.

*2.7 Model validation*

Computational results were validated with Doppler ultrasound data. Fig. 3 shows the percentage of blood flow at each outlet of the aorta for Doppler ultrasound measurement and numerical simulation prediction. The peak of the measured waveform is shifted to the peak systole of the inlet blood flow waveform, which is indicated by the vertical dashed line. Results show the rigid wall assumption fails to capture the true waveform and predicts non-existent backflow. For the two FSI scenarios, good agreements on the waveform and magnitude are observed at the three outlets of the aortic branches, while the biggest deviation occurs in the left common carotid artery. FSI method would improve the prediction results of our computational framework and the endograft has a slight effect on the blood flow distribution.

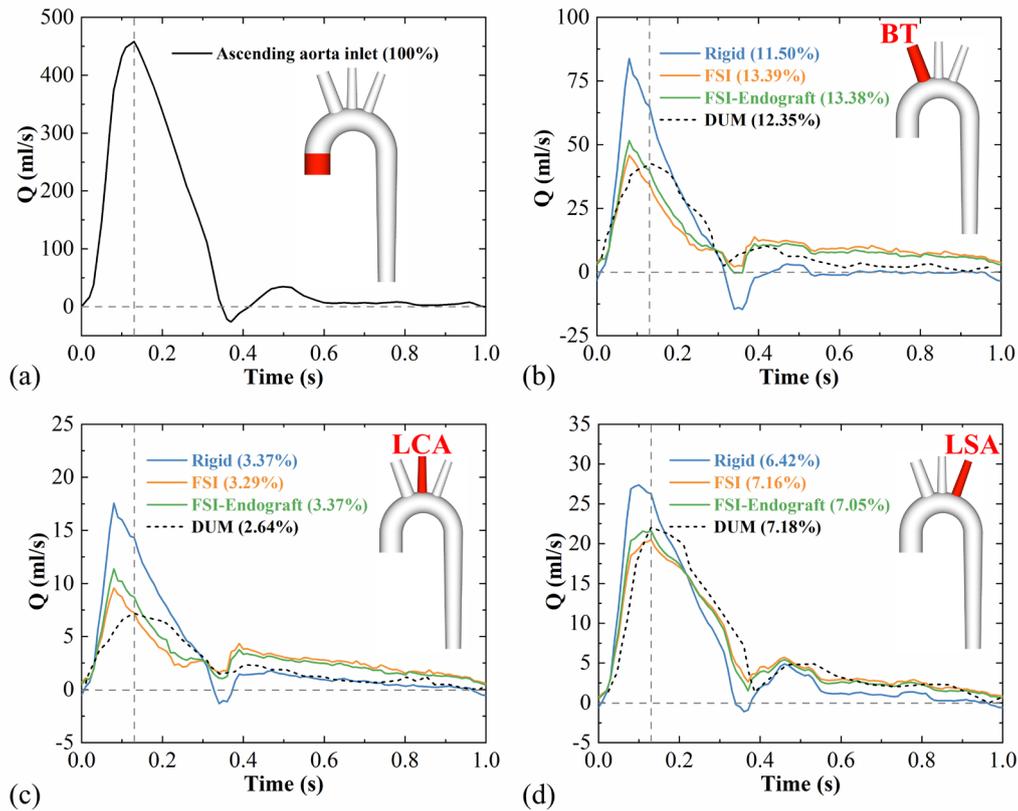

**Fig. 3.** Model validation. (a) Inlet boundary condition of ascending aorta: a normalized physiological flow waveform. Percentage of blood flow at each outlet of the aorta for Doppler ultrasound measurement (DUM) and numerical simulation prediction: (b) BT: brachiocephalic artery; (c) LCA: left carotid artery; (d) LSA: left subclavian artery. The vertical dashed line indicates the peak systole.

## 3. Results

### *3.1 Energy loss and aortic volume*

Human aortas have a buffer function. The proximal aorta expands in volume and stores energy during systole, which is released through a contraction in the diastole. The comparison of EL variations during a cardiac cycle is shown in Fig. 4. The gray background area indicates the systolic phase, which includes two stages of blood acceleration and deceleration. In Fig. 4a, the EL continues to increase at the beginning and reaches the maximum before peak systole. For rigid wall assumption, the EL decreases to the negative minimum and then elevates to zero when the systolic phase ends. At the same time, the negative minimum EL of FSI scenarios is observed and the corresponding value is relatively lower. It should be noted that the EL keeps negative during the diastole in the two FSI scenarios. The three horizontal dashed lines indicate the mean EL during a cardiac cycle and the corresponding values are marked nearby. The mean EL of the FSI scenarios is significantly higher than that of the rigid wall (3.34 times). When the effect of the endograft is considered, the magnitude of EL decreases, and the mean value receives a reduction of 19.1%.

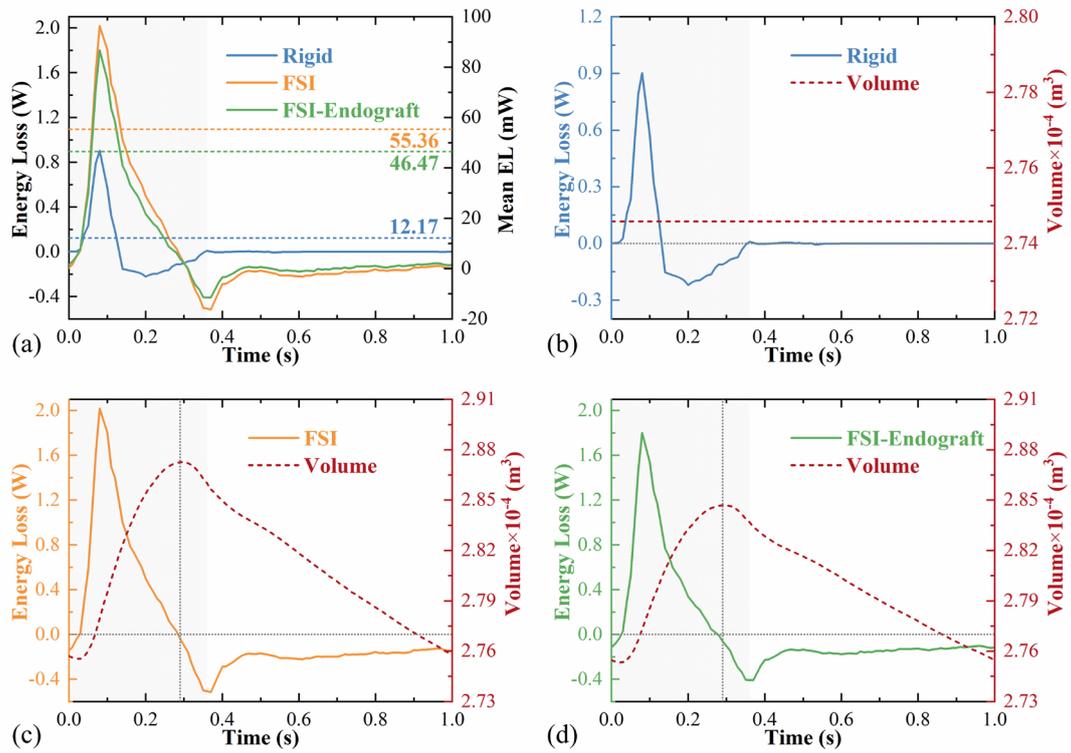

**Fig. 4.** (a) Energy loss (EL) during a cardiac cycle. The gray background area indicates the systole. Three horizontal dashed lines indicate mean EL and the corresponding values are marked. (b) Rigid wall. There is no change in aortic volume. (c) Fluid-structure interaction. The vertical dotted line represents a time point when the maximum volume of the aorta and zero EL is observed. (d) Fluid-structure interaction with endograft.

Changes in the aortic volume are also investigated in Fig. 4. There is no change in the volume of the aorta using the rigid wall assumption (Fig. 4b). For FSI scenarios, the aorta contracts slightly at the beginning, and a phase difference between the aortic volume and EL is observed (Fig. 4c, d). As the aortic volume continues to expand, the EL first increases to the peak value and then drops to zero. At this time point, the maximum volume of the aorta is observed, which is depicted by a vertical dotted line. It is worth emphasizing that the presence of the endograft limits the deformability of the aorta.

Fig. 5 further explores the relationship between the EL and aortic volume in FSI scenarios. Four phases are observed during a cardiac cycle. First, EL increases to a maximum with volume expansion. Second, EL decreases to zero with volume expansion. Third, EL decreases to a minimum with volume shrinkage. Fourth, EL increases to zero with volume shrinkage. The aorta stores blood flow energy by expanding during the first and second phases, while part of the energy is released through volume shrinkage in the third and fourth phases.

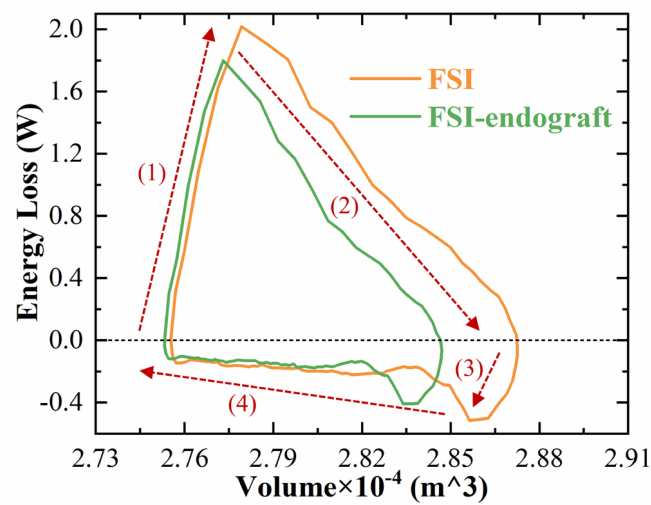

**Fig. 5.** Relationship between energy loss (EL) and aortic volume for two FSI scenarios. Four

phases are observed during a cardiac cycle. First, EL increases to a maximum with volume expansion. Second, EL decreases to zero with volume expansion. Third, EL decreases to a minimum with volume shrinkage. Fourth, EL increases to zero with volume shrinkage.

### *3.2 Wall pressure and aortic volume*

Wall pressure exerted by the blood flow is also associated with EL. Pulsation waveforms of average wall pressure and aortic volume during a cardiac cycle are illustrated in Fig. 6. The magnitudes of the pressure change agree well with the clinical measurement data. A transient dip can be observed at the end of the systole, which represents the closure of the aortic valve. Fig. 6c shows the relationship between pressure and volume. When the aorta is expanding, the area enclosed by the corresponding curve and the horizontal axis means the total work ($W_t$) done by the blood on the aortic wall. When the aorta shrinks, the dilated wall would conversely do work on the blood, which can also be calculated by the corresponding area. The difference between the above two areas is exactly equal to the area of the closed curve, which represents the EL related to the deformation of the aortic wall and accounts for more than 20% of the total work ($W_t$). The closed curve can be divided into four phases (Fig. 6d). First, pressure increases with volume shrinkage. Second, pressure increases to a maximum with volume expansion. Third, pressure decreases with volume expansion. Fourth, pressure decreases to a minimum with volume shrinkage. The distribution of blood flow EL is shown in Table 2. For the rigid wall, all EL results from blood flow, while both the deformation of the aortic wall and flow loss contribute to EL in FSI scenarios. The corresponding formula of mean EL is as follows:

$$EL_{rigid} = EL_{flow} \tag{9}$$

$$EL_{FSI} = EL_{flow} + EL_{P-V} \tag{10}$$

It should be emphasized that EL caused by the interaction between the blood flow and the aortic wall accounts for 68.8% of the total EL.

TABLE 2. The calculation results of EL during a cardiac cycle.

| Case | Mean EL | EL$_{P-V}$ | EL$_{flow}$ |
|---|---|---|---|
| **Rigid** | 12.17 | \ | 12.17 |
| **FSI** | 55.36 | 40.54 (73.2%) | 14.82 (26.8%) |
| **FSI-endograft** | 46.47 | 31.97 (68.8%) | 14.50 (31.2%) |

The mean EL is the sum of EL caused by the blood-wall interaction (EL$_{P-V}$) and blood flow (EL$_{flow}$). In FSI scenarios, EL$_{P-V}$ is calculated through the area of the P-V diagram (Unit: mW).

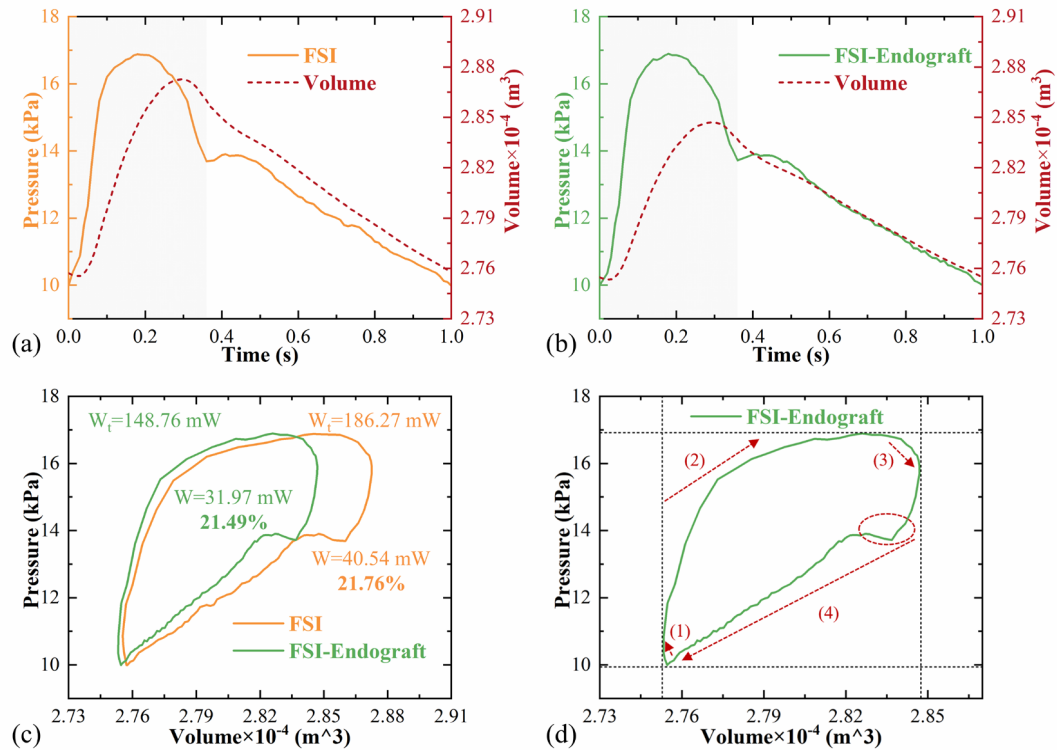

**Fig. 6.** Average wall pressure waveform during a cardiac cycle. (a) Fluid-structure interaction (FSI). (b) FSI with endograft. The gray background area indicates the systole. (c) The relationship between EL and aortic volume for two FSI scenarios. (d) Four phases. First, pressure increases with volume shrinkage. Second, pressure increases to a maximum with volume expansion. Third, pressure decreases with volume expansion. Fourth, pressure decreases to a minimum with volume shrinkage. The dotted lines indicate extremums of wall pressure and aortic volume.

### 3.3 Wall shear stress-related indices

The distribution of the WSS-related indices is illustrated in Fig. 7. Regions exposed to low TAWSS (< 0.4 Pa) and high OSI (> 0.25) are analyzed, which could identify the risk of atherosclerosis and in-stent thrombus formation. There is no significant change in the risk area when different wall simulation strategies are applied. Only a slight difference can be observed on the three branches of the aortic arch. In the FSI scenario with endograft, low TAWSS regions appear at the ascending aorta and two nearby branches, which coincides with high OSI areas. It should be noted that these regions are at high risk of atherosclerosis.

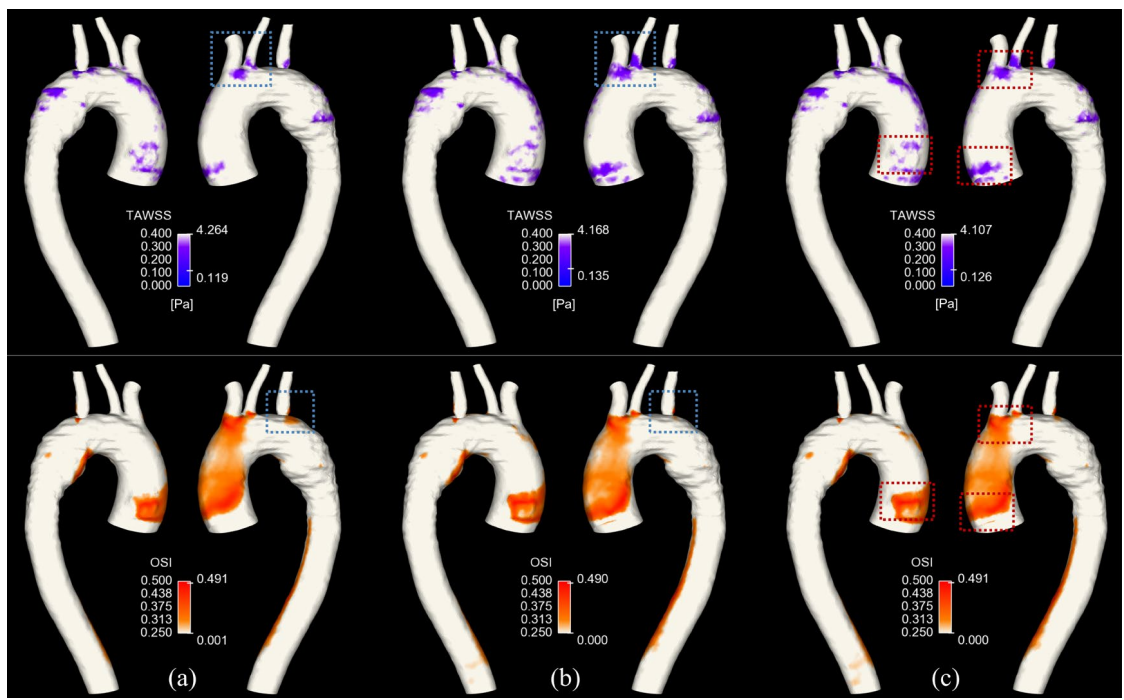

**Fig. 7.** Wall shear stress-related indices. (a) Rigid wall. (b) Fluid-structure interaction (FSI). (c) FSI with endograft. Top: The comparison of low time-averaged wall shear stress (< 0.4 Pa) distribution. Bottom: The comparison of high oscillatory shear index (> 0.25) distribution. The blue dotted boxes indicate the area of difference. The red dotted boxes denote high-risk areas. The color axes are all set to the same range for comparison and the extreme values are shown on the right side of the color axes.

*3.4 Aortic wall deformation*

Fig. 8 shows the comparison of maximum aortic wall deformation (t=0.29 s). The time point is on the eve of the end of the systole. The largest deformed region locates in the ascending aorta and aortic arch, while the displacement of three arch branches can be neglected. When the effect of the endograft is considered, the composite stent-wall region was not easily deformed relative to the aortic wall. Specifically, the deformations of the aortic arch and proximal descending aorta reduce obviously. A relatively high deformed region is captured at the descending aorta adjacent to the distal stent-wall region.

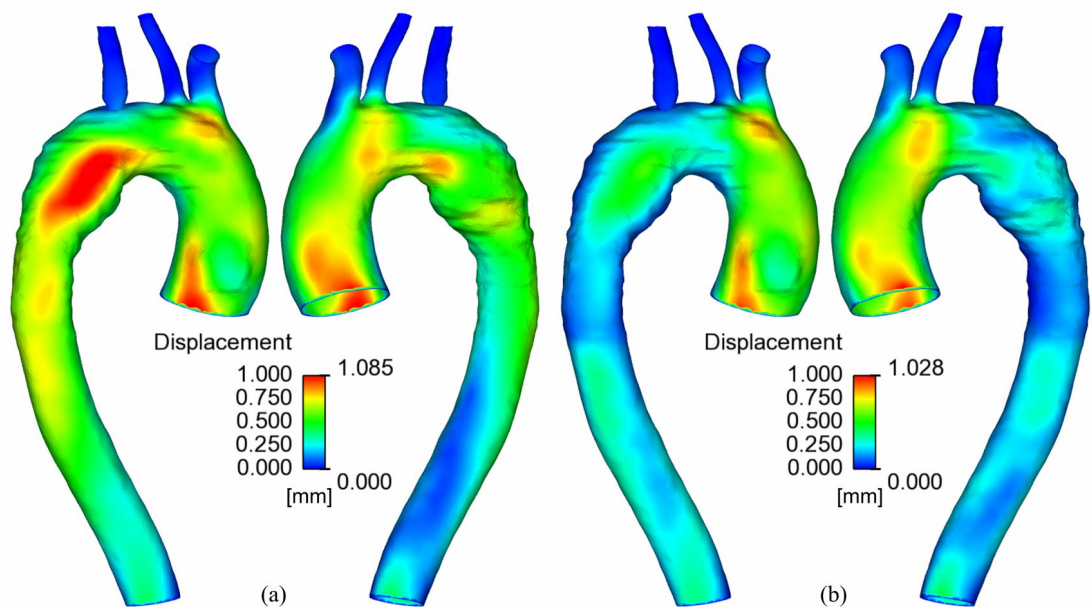

**Fig. 8.** Comparison between maximum aortic wall deformation at t=0.29 s. The time point is on the eve of the end of the systole. (a) Fluid-structure interaction (FSI). (b) FSI with endograft. The color axes are all set to the same range for comparison and the extreme values are shown on the right side of the color axes.

## 4. Discussion

The interest in biomechanical implications of the endograft after TEVAR has been increasing and a strong relationship between the postoperative complications and implanted endograft was reported in recent clinical studies. van Bakel et al. [27] found that the stiffness mismatch between the endograft and aortic wall may be responsible for the increased left ventricular stroke work and mass during follow-up. Zhu et al. [28] demonstrated that aortic flow patterns and WSS distribution were significantly altered by a double-branched endograft which is consistent with our previous finding [29]. The present study was designed to determine the biomechanical implications of the endograft by the advanced FSI method.

Rigid wall assumption only considers the EL caused by blood flow, while the part related to wall deformation can also be captured by the FSI method. During systole, negative EL in the rigid wall case is observed, which is similar to the energy release process from the perspective of EL. However, this phenomenon is a result of pulsating blood flow. Specifically, when the inlet blood flow slows down, the high-speed blood flow that entered previously has reached the outlets, therefore the EL at this time is negative. It is more appropriate to call EL apparent energy loss.

EL does not mean that the energy is consumed, part of it is just stored temporarily and would be released later in FSI scenarios. The most interesting finding is that the complete storage and release process of blood flow energy could be captured by our FSI method, which consists of four phases. It should be noted that aortic volume is the most intuitive indicator of EL. When aortic volume is expanding, EL would keep a positive value. Until the aortic volume begins to contract, EL becomes negative. Besides, Young's modulus of the endograft is relatively higher than that of the aortic wall, resulting in an elevated stiffness of the composite stent-wall region. Previous studies have reported that the increased stiffness of the postoperative aorta is related to the inner mechanical behavior of the endograft and the percentage of the aorta covered by endograft [30, 31]. Therefore, the deformability of the aortic wall is limited by the implantation of the endograft, which also weakens the aortic buffer function and reduces the magnitude of mean EL.

Mean EL can measure the real EL caused by the blood internal collision and interaction between blood flow and the aortic wall. Our results show that the latter is responsible for the majority of EL (68.8%) during a cardiac cycle. The diagram of wall pressure and aortic volume is applied to determine the mutual work between the blood and the aortic wall. Additionally, the area of the closed curve can be used to calculate the EL caused by the aortic wall deformation ($EL_{P-V}$). When the endograft is implanted, $EL_{P-V}$ significantly reduces (26.8%), while the change of $EL_{flow}$ is negligible, indicating that the EL associated with blood flow is insensitive to the endograft. In summary, the composition of EL can be determined by our advanced FSI method with endograft.

A previous study has revealed that regions of low-TAWSS and high-OSI of aortic dissection can only be identified accurately in the FSI simulation [22]. We confirmed that OSI distributions in the FSI model and the rigid wall model show a significant difference in a patient with Stanford type B dissection [32]. However, we found that FSI and endograft have no obvious effect on the WSS-related-indices based on the analysis of low-TAWSS and high-OSI distributions in this paper, where the patient suffered from an aortic arch aneurysm. Considering the complex geometric model of aortic dissection in previous FSI studies, the false lumen may be responsible for the remarkable difference. Therefore, we suggest the rigid wall assumption is suitable for the analysis of WSS-related-indices in patients with aortic aneurysms when computing resources are limited.

We found that the maximum aortic wall deformation appeared at t=0.29 s, which is different from the peak systole (t=0.13 s). The implantation of the endograft greatly restricts the deformation of the aortic arch and proximal descending aorta. Moreover, the above areas are located in the composite stent-wall region. It should be emphasized that the stiffness of the stent-wall region (10.4 MPa) is more than five times that of the aortic wall (2 MPa). There is a clear boundary for wall displacement in the middle of the descending aorta, which might result from the mismatch between the endograft and the aortic wall.

In this paper, a novel image-based, patient-specific, fluid-structure computational framework was proposed to offer new insights into the biomechanical compilations of the endograft, but there are still some limitations that need to be emphasized. First, only one patient treated by TEVAR was analyzed. It should be noted that the present study is a pilot report aimed at presenting our computational model and the aortic geometry is representative from the perspective of the clinic. More patient-specific postoperative cases will need to further evaluate the effect of the endograft. Second, the thickness of the endograft and aortic wall were simplified to be uniform. Oliveira et al. [17] generated the non-uniform aortic wall and the thickness is proportional to the radius of the lumen. This strategy would be adopted in our follow-up study. The implanted endograft consists of a textile membrane and a metal structure, we would reconstruct the real geometric model of the endograft according to CTA image and *in vitro* measurement data. Third, the linear elastic property was applied to the aortic wall in the present study. Compared with the linear elastic assumption, the hyperelastic model can more accurately describe the mechanical characteristics of the aortic wall. Torii et al.

[33] found that there are no remarkable differences between the velocity fields predicted by a hyperelastic and a linear elastic model for the arterial wall in FSI simulations. Therefore, the linear elastic assumption is a reasonable compromise between the accuracy and technical difficulties. Forth, a flat flowrate profile was applied as the inflow boundary condition. A recent study has pointed out that fully patient-specific inlet velocity profiles should be used to produce meaningful results [34]. 4D PC-MRI measurement data would be adopted to extract the patient-specific velocity profiles in our future study. Finally, FSI is time-consuming relative to rigid wall, which limits the applicability in clinical. GPU acceleration and algorithm optimization in the future may solve this problem.

## 5. Conclusions

This study evaluates the biomechanical implications of the endograft after TEVAR. Our results demonstrate that FSI can not only acquire the deformation of the aortic wall but also capture the complete storage and release process of blood flow energy during a cardiac cycle, where the magnitude of EL would be overpredicted if the endograft is neglected (19.1%). The relationship between the EL and aortic volume is firstly constructed and the process of EL change is divided into four phases. The second major finding is that the closed curve area of wall pressure and aortic volume can indicate EL caused by the interaction between the blood flow and the aortic wall, which accounts for 68.8% of the total EL. Besides, there is a slight difference in the distribution of WSS-related-indices between the three scenarios. Finally, we found that the presence of the endograft has a dramatic effect on the postoperative deformation of the aortic wall. In conclusion, the interaction between blood flow, the implanted endograft, and the aortic wall should be considered in the post-TEVAR computational hemodynamic study, which is conducive to acquire physiologically accurate hemodynamics.

**Acknowledgments**

This research was supported by the National Postdoctoral Program for Innovative Talents (CN) [grant number BX20200290], Postdoctoral Science Foundation (CN) [grant number 2020M681852], Postdoctoral Science Foundation of Zhejiang Province (CN) [grant number ZJ2020153].

**Conflict of interest**

All authors declare that they have no conflicts of interest.

**Ethical approval**

This study was approved by the Ethics Committee of Zhongshan Hospital, Fudan University, Shanghai, China (Ref No. Y2015-193).

**REFERENCES**
[1] A. Harky, J.S.K. Chan, C.H.M. Wong, N. Francis, C. Grafton-Clarke, M. Bashir, Systematic review and meta-analysis of acute type B thoracic aortic dissection, open, or endovascular repair, J. Vasc. Surg., 69 (2019) 1599-1609 e1592, https://doi.org/10.1016/j.jvs.2018.08.187.
[2] M. Alsawas, F. Zaiem, L. Larrea-Mantilla, J. Almasri, P.J. Erwin, G.R. Upchurch, Jr., M.H. Murad, Effectiveness of surgical interventions for thoracic aortic aneurysms: A systematic review and meta-analysis, J. Vasc. Surg., 66 (2017) 1258-1268 e1258, https://doi.org/10.1016/j.jvs.2017.05.082.
[3] R.O. Afifi, A. Azizzadeh, A.L. Estrera, Complications of TEVAR, Surgical Management of Aortic Pathology, Springer2019, pp. 1211-1223.
[4] Y. Kanaoka, T. Ohki, K. Maeda, T. Baba, Analysis of risk factors for early type I endoleaks after thoracic endovascular aneurysm repair, J. Endovasc. Ther., 24 (2017) 89-96,

https://doi.org/10.1177/1526602816673326.

[5] Z.H. Dong, W.G. Fu, Y.Q. Wang, D.Q. Guo, X. Xu, Y. Ji, B. Chen, J.H. Jiang, J. Yang, Z.Y. Shi, T. Zhu, Y. Shi, Retrograde type A aortic dissection after endovascular stent graft placement for treatment of type B dissection, Circulation, 119 (2009) 735-741, https://doi.org/10.1161/CIRCULATIONAHA.107.759076.

[6] Z. Dong, W. Fu, Y. Wang, C. Wang, Z. Yan, D. Guo, X. Xu, B. Chen, Stent graft-induced new entry after endovascular repair for Stanford type B aortic dissection, J. Vasc. Surg., 52 (2010) 1450-1457, https://doi.org/10.1016/j.jvs.2010.05.121.

[7] D. Daye, T.G. Walker, Complications of endovascular aneurysm repair of the thoracic and abdominal aorta: evaluation and management, Cardiovasc Diagn Ther, 8 (2018) S138-S156, https://doi.org/10.21037/cdt.2017.09.17.

[8] P. Geisbusch, D. Skrypnik, M. Ante, M. Trojan, T. Bruckner, F. Rengier, D. Bockler, Endograft migration after thoracic endovascular aortic repair, J. Vasc. Surg., 69 (2019) 1387-1394, https://doi.org/10.1016/j.jvs.2018.07.073.

[9] C.A. Figueroa, C.A. Taylor, A.J. Chiou, V. Yeh, C.K. Zarins, Magnitude and direction of pulsatile displacement forces acting on thoracic aortic endografts, J. Endovasc. Ther., 16 (2009) 350-358, https://doi.org/10.1583/09-2738.1.

[10] A. Prasad, N. Xiao, X.Y. Gong, C.K. Zarins, C.A. Figueroa, A computational framework for investigating the positional stability of aortic endografts, Biomech Model Mechanobiol, 12 (2013) 869-887, https://doi.org/10.1007/s10237-012-0450-3.

[11] C. Menichini, S. Pirola, B. Guo, W. Fu, Z. Dong, X.Y. Xu, High wall stress may predict the formation of stent-graft–induced new entries after thoracic endovascular aortic repair, J. Endovasc. Ther., 25 (2018) 571-577, https://doi.org/10.1177/1526602818791827.

[12] T.M.J. van Bakel, N.S. Burris, H.J. Patel, C.A. Figueroa, Ascending aortic rupture after zone 2 endovascular repair: a multiparametric computational analysis, Eur. J. Cardiothorac. Surg., 56 (2019) 618-621, https://doi.org/10.1093/ejcts/ezy458.

[13] Y. Qiao, L. Mao, T. Zhu, J. Fan, K. Luo, Biomechanical implications of the fenestration structure after thoracic endovascular aortic repair, J. Biomech., 99 (2019) 109478, https://doi.org/10.1016/j.jbiomech.2019.109478.

[14] L. Wang, X. Zhou, D. Guo, K. Hou, Z. Shi, X. Tang, W. Fu, A New Adjustable Puncture Device for In Situ Fenestration During Thoracic Endovascular Aortic Repair, J. Endovasc. Ther., (2018) 1526602818776623, https://doi.org/10.1177/1526602818776623.

[15] F. Gijsen, E. Allanic, F. Van de Vosse, J. Janssen, The influence of the non-Newtonian properties of blood on the flow in large arteries: unsteady flow in a 90 curved tube, J. Biomech., 32 (1999) 705-713, https://doi.org/10.1016/S0021-9290(99)00014-7.

[16] S. Pasta, J.S. Cho, O. Dur, K. Pekkan, D.A. Vorp, Computer modeling for the prediction of thoracic aortic stent graft collapse, J. Vasc. Surg., 57 (2013) 1353-1361, https://doi.org/10.1016/j.jvs.2012.09.063.

[17] D. Oliveira, S.A. Rosa, J. Tiago, R.C. Ferreira, A.F. Agapito, A. Sequeira, Bicuspid aortic valve aortopathies: An hemodynamics characterization in dilated aortas, Comput. Methods Biomech. Biomed. Eng., 22 (2019) 815-826, https://doi.org/10.1080/10255842.2019.1597860.

[18] E. Di Martino, G. Guadagni, A. Fumero, G. Ballerini, R. Spirito, P. Biglioli, A. Redaelli, Fluid–structure interaction within realistic three-dimensional models of the aneurysmatic aorta as a guidance to assess the risk of rupture of the aneurysm, Med. Eng. Phys., 23 (2001) 647-655, https://doi.org/10.1016/S1350-4533(01)00093-5.

[19] J. Alastruey, N. Xiao, H. Fok, T. Schaeffter, C.A. Figueroa, On the impact of modelling assumptions in multi-scale, subject-specific models of aortic haemodynamics, J R Soc Interface, 13 (2016), https://doi.org/10.1098/rsif.2016.0073.

[20] R. Nerem, W. Seed, N. Wood, An experimental study of the velocity distribution and transition to turbulence in the aorta, J. Fluid Mech., 52 (1972) 137-160, https://doi.org/10.1017/S0022112072003003.

[21] S. Pirola, Z. Cheng, O.A. Jarral, D.P. O'Regan, J.R. Pepper, T. Athanasiou, X.Y. Xu, On the choice of outlet boundary conditions for patient-specific analysis of aortic flow using computational fluid dynamics, J. Biomech., 60 (2017) 15-21, https://doi.org/10.1016/j.jbiomech.2017.06.005.

[22] M. Alimohammadi, J.M. Sherwood, M. Karimpour, O. Agu, S. Balabani, V. Diaz-Zuccarini, Aortic dissection simulation models for clinical support: fluid-structure interaction vs. rigid wall models, Biomed Eng Online, 14 (2015) 34, https://doi.org/10.1186/s12938-015-0032-6.

[23] A.G. Brown, Y. Shi, A. Marzo, C. Staicu, I. Valverde, P. Beerbaum, P.V. Lawford, D.R. Hose, Accuracy vs. computational time: translating aortic simulations to the clinic, J. Biomech., 45 (2012) 516-523, https://doi.org/10.1016/j.jbiomech.2011.11.041.

[24] R. Campobasso, F. Condemi, M. Viallon, P. Croisille, S. Campisi, S. Avril, Evaluation of Peak Wall Stress in an Ascending Thoracic Aortic Aneurysm Using FSI Simulations: Effects of Aortic Stiffness and Peripheral Resistance, Cardiovasc. Eng. Technol., 9 (2018) 707-722, https://doi.org/10.1007/s13239-018-00385-z.

[25] T.E. Tezduyar, S. Sathe, T. Cragin, B. Nanna, B.S. Conklin, J. Pausewang, M. Schwaab, Modelling of fluid–structure interactions with the space–time finite elements: Arterial fluid mechanics, International Journal for Numerical Methods in Fluids, 54 (2007) 901-922, https://doi.org/10.1002/fld.1443.

[26] K. Itatani, K. Miyaji, Y. Qian, J.L. Liu, T. Miyakoshi, A. Murakami, M. Ono, M. Umezu, Influence of surgical arch reconstruction methods on single ventricle workload in the Norwood procedure, J. Thorac. Cardiovasc. Surg., 144 (2012) 130-138, https://doi.org/10.1016/j.jtcvs.2011.08.013.

[27] T.M.J. van Bakel, C.J. Arthurs, F.J.H. Nauta, K.A. Eagle, J.A. van Herwaarden, F.L. Moll, S. Trimarchi, H.J. Patel, C.A. Figueroa, Cardiac remodelling following thoracic endovascular aortic repair for descending aortic aneurysms, Eur. J. Cardiothorac. Surg., 55 (2019) 1061-1070, https://doi.org/10.1093/ejcts/ezy399.

[28] Y. Zhu, W. Zhan, M. Hamady, X.Y. Xu, A pilot study of aortic hemodynamics before and after thoracic endovascular repair with a double-branched endograft, Medicine in Novel Technology and Devices, 4 (2019), https://doi.org/10.1016/j.medntd.2020.100027.

[29] Y. Qiao, L. Mao, Y. Ding, J. Fan, T. Zhu, K. Luo, Hemodynamic consequences of TEVAR with in situ double fenestrations of left carotid artery and left subclavian artery, Med. Eng. Phys., 76 (2020) 32-39, https://doi.org/10.1016/j.medengphy.2019.10.016.

[30] M. Domanin, G. Piazzoli, S. Trimarchi, C. Vergara, Image-Based Displacements Analysis and Computational Blood Dynamics after Endovascular Aneurysm Repair, Ann. Vasc. Surg., 69 (2020) 400-412, https://doi.org/10.1016/j.avsg.2020.07.014.

[31] H.W. De Beaufort, M. Conti, A.V. Kamman, F.J. Nauta, E. Lanzarone, F.L. Moll, J.A. Van Herwaarden, F. Auricchio, S. Trimarchi, Stent-graft deployment increases aortic stiffness in an ex vivo porcine model, Ann. Vasc. Surg., 43 (2017) 302-308, https://doi.org/10.1016/j.avsg.2017.04.024.

[32] Y. Qiao, Y. Zeng, Y. Ding, J. Fan, K. Luo, T. Zhu, Numerical simulation of two-phase non-Newtonian blood flow with fluid-structure interaction in aortic dissection, Comput. Methods Biomech. Biomed. Eng., 22 (2019) 620-630, https://doi.org/10.1080/10255842.2019.1577398.

[33] R. Torii, M. Oshima, T. Kobayashi, K. Takagi, T.E. Tezduyar, Fluid–structure interaction modeling

of a patient-specific cerebral aneurysm: influence of structural modeling, Computational Mechanics, 43 (2008) 151-159, https://doi.org/10.1007/s00466-008-0325-8.

[34] P. Youssefi, A. Gomez, C. Arthurs, R. Sharma, M. Jahangiri, C. Alberto Figueroa, Impact of Patient-Specific Inflow Velocity Profile on Hemodynamics of the Thoracic Aorta, J. Biomech. Eng., 140 (2018), https://doi.org/10.1115/1.4037857.